\documentclass{ws-mpla}
\usepackage{graphicx,psfrag}
\usepackage{psfig}
\begin{document}

\markboth{R Tharanath and V C Kuriakose}
{Thermodynamics and Spectroscopy of Schwarzschild black hole surrounded by quintessence}

\catchline{}{}{}{}{}

\title{THERMODYNAMICS AND SPECTROSCOPY OF SCHWARZSCHILD BLACK HOLE SURROUNDED BY QUINTESSENCE}

\author{\footnotesize  R Tharanath\footnote{
Typeset names in 8 pt Times Roman, uppercase. Use the footnote to 
indicate the present or permanent address of the author.}}

\address{Department of Physics, Cochin University of Science and
Technology, Kochi 682022, India
tharanath.r@gmail.com}

\author{V C Kuriakose}

\address{Department of Physics, Cochin University of Science and
Technology, Kochi 682022, India\\
vck@cusat.ac.in
}

\maketitle

\pub{Received (Day Month Year)}{Revised (Day Month Year)}

\begin{abstract}
 The thermodynamic and spectroscopic behaviour of Schwarzschild black hole surrounded by quintessence are studied.
We have derived the thermodynamic quantities and studied their behaviour for different values of quintessence parameter.
 We put the
background space-time into the Kruskal-like coordinate to find the period with respect to
Elucidean time. Also assuming that the adiabatic invariant obeys Bohr-Sommerfeld quantization rule, detailed study of 
area spectrum and entropy spectrum have been done for special cases of the quintessece state parameter.
 We find that the spectra are equally spaced.

\end{abstract}

\ccode{PACS Nos.: 04.70.Dy, 04.70.-s }

\section{Introduction}	
Accelerating expansion of the universe is one of the most recent fascinating results of observational cosmology. 
To explain the accelerated expansion of the universe, it is proposed that the universe is regarded as being dominated
by an exotic scalar field with a large negative pressure called ``dark energy'' which constitutes about 
70 percent of the total energy of the universe. There are several candidates for dark energy. ``Quintessence''\cite{Caldwell,Freese} is one among them.
It is characterized by a parameter $\epsilon$, the ratio
of the pressure to energy density of the dark energy, and the value of $\epsilon$ falls in the range $-1\leq\epsilon\leq-\frac{1}{3}$. 
 The quasi normal modes of different black holes surrounded
by quintessence have been studied earlier\cite{Songhai,Nijo}.
We now investigate the effect of quintessence on the area spectrum of a Schwarzschild black hole. 
 It would be also interesting to know how does the dark energy component affect the thermodynamics of black holes

The quantization of the black hole horizon area is one of the most interesting subjects of quantum gravity
ever since Bekenstein first proposed in 1974 that the black hole area spectrum is equally spaced\cite{Bekenstein}. Bekenstein has conjectured the possibility of a 
connection between quasi normal modes of black holes, which have been studied extensively, and the area spectrum.

 It is widely believed that the area eigenvalues are equally spaced, $A=\gamma l_{Pl}^{2}$, where $\gamma$ is a dimensionless 
constant and $l_{Pl}$ is the Planck length. However there is no general agreement on the value of the spacing factor  $\gamma$. 
Bekenstein was the first physicist, who introduced the notion that, when a neutral particle is absorbed by a black hole, the horizon area is
 increased by $\gamma =8 \pi$. Later many methods of calculations have brought the same result. 
These examples include the Maggiore's reinterpretation
\cite{Maggiore} of the renowned Hod conjecture \cite{Hod}, a quantization procedure proposed by Ropotenko
\cite{ropo}, a refinement there of \cite{med}, a method reducing the black hole phase space to a pair
of observables \cite{bra}, identifying the exponent of the gravitational action as a quantum
amplitude \cite{Padmanabhan} and a recent attempt
to use the adiabatic invariance \cite{majhi}, etc. On the other hand many have questioned the spacing parameter in
the context of quasinormal modes. They have argued that the spacing factor would be $\gamma= 4 ln 3$ \cite{Hod,Setare,Kunstatter}.
And a recent observation by Banerjee, Majhi and Vagenas \cite{ban}has shown that the area spacing parameter of a black hole horizon is fixed by $\gamma= 4$.
 There is no general agreement on the value of area spacing parameter, here we are going to probe the effects of dark energy 
components in the value of this spacing factor. 

In the present work, the calculation employs
the proposal of Zeng et al\cite{zeng} to probe the values of the 
area spacing parameter for Schwarzschild black hole surrounded by quintessence. Since the gravity system is periodic with respect to
Euclidean time with a period given by the inverse of the Hawking temperature, it
is assumed that the frequency of the outgoing wave is given by this temperature.
So the adiabatic invariant, which in turn obeys the Bohr's correspondence principle, will contain the surface gravity term.
 In the method of using the asymptotic quasinormal modes to find the area spectrum, 
the surface gravity term in the adiabatic invariant appears to be the damping part of the quasinormal modes. In short, to
find the area spectrum we should find the adiabatic invariant integral which includes
the surface gravity for particular horizons. In the present study of Schwarzschild black hole 
surrounded by quintesence, we obtain different horizons for different values of quintessence parameter. Thus, this can be taken as
a generalized study of the effect of dark energy components in the area and entropy spectra. 

In section 2 we study the thermodynamics and phase transition of  the Schwarzschild black hole surrounded by quintessence. 
 We calculate the area and entropy spectrum of the system for three different cases 
of quintessence parameter $\epsilon$ in section 3. And the results and conclusion of the present study are summarized
in the section 4. 
\section{Thermodynamics of Schwarzschild black hole surrounded by Quintessence}
If the black hole is regarded as
a thermal system, it is then natural to apply the laws of thermodynamics; however,
a crucial difference from the other thermal systems is that it is a gravitational object whose
entropy is identified with the area of the black hole(In this section we are using $c=G=\hbar=1$).
The metric of Schwarzschild black hole surrounded by quintessence\cite{kiselev} is given by,
\begin{equation}\label{metric}
  ds^{2}=f(r)dt^{2}-\frac{1}{f(r)} dr^{2}-r^{2}(d\theta^{2}+\sin\theta^{2}d\phi^{2}),
\end{equation}
where
 \begin{equation}\label{f(r)}
  f(r)=1-\frac{2M}{r}-\frac{a}{r^{3\epsilon+1}}.
\end{equation}
Here M is the black hole mass and $a$ is the normalization factor, which is positive, depending on the energy density
 of quintessence. Quintessence is a scalar field whose equation of state parameter $\epsilon$ is defined as the ratio of its pressure $p$ and its
energy density $\rho$, which is given by a kinetic term and a potential term as\cite{Caldwell}, $\epsilon\equiv\frac{p}{\rho}=\frac{\frac{1}{2}\dot Q^{2}-V(Q)}{\frac{1}{2}\dot Q^{2}+V(Q)}$.
Following Kiselev\cite{kiselev} the energy density can be written as $\rho_{\epsilon}=-\frac{a}{2}\frac{3\epsilon}{r^{3(1+\epsilon)}}$.

            We can establish the relation between mass of a black hole and its horizon radius directly from the above
equation as, 
\begin{equation}\label{mr}
 M=\frac{r}{2}-\frac{a}{2 r^{3\epsilon}},
\end{equation}
and we know that entropy can be written as
\begin{equation}\label{sa}
 S=\frac{A}{4}=\frac{4 \pi r^{2}}{4}= \pi r^{2},
\end{equation}
so that $r$ can be written in terms of $S$ as 
\begin{equation}\label{rs}
 r= \sqrt{\frac{S}{\pi}}.
\end{equation}
Let us rewrite (\ref{mr}) as
\begin{equation}\label{ms}
 M=\frac{1}{2}\left[\sqrt{\frac{S}{\pi}} - a(\frac{\pi}{S})^{\frac{3 \epsilon}{2}}\right].
\end{equation}
In Fig(1) we have plotted the variation of mass of black hole with respect to entropy. Mass increases as the entropy  increases,
and it is evident that the horizon area also increases. Since we have an area law of entropy, the increase in area will 
cause the increase in entropy.
\begin{figure}
 \centering
\includegraphics[width=0.60\columnwidth]{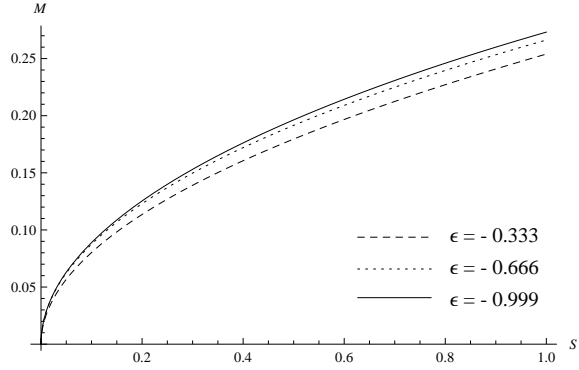}
\caption{Variation of mass with entropy for different values of $\epsilon$, keeping $a=0.001$.}
\end{figure}

       The density of quintessence $\rho_{\epsilon}$ can be written as 
\begin{equation}\label{ror}
\rho_{\epsilon}=-\frac{a}{2}\frac{3\epsilon}{r^{3(1+\epsilon)}}.
\end{equation}
In terms of entropy $S$ the above equation will be
\begin{equation}\label{ros}
 \rho_{\epsilon}=- \frac{3 a \epsilon}{2}(\frac{\pi}{S})^{\frac{3 \epsilon +3}{2}}.
\end{equation}
The variation of the density parameter with the entropy is plotted for different values of $a$ in Fig(2).
 It shows that density factor decreases as entropy increases. Its variation is almost same for all
values of the quintessence parameter.
\begin{figure}
 \centering
\includegraphics[width=0.60\columnwidth]{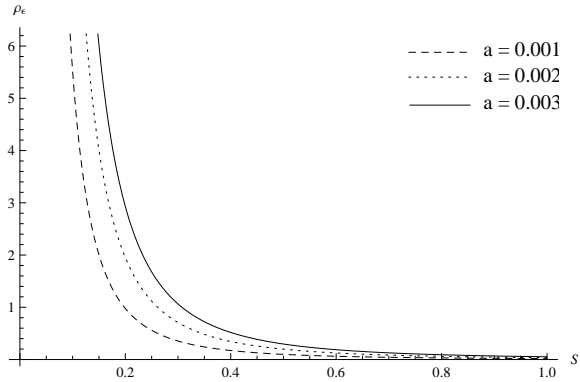}
\caption{Variation of density of quintessence with entropy for different values of $a$, keeping $\epsilon=-\frac{2}{3}$.}
\end{figure}
Now we can deduce the thermodynamical quantities from the above expression of mass in terms of entropy.

 \begin{subequations}
  \begin{equation}\label{t}
   T=\left(\frac{\partial M}{\partial S}\right),
  \end{equation}
\begin{equation}\label{c}
 C=T\left(\frac{\partial S}{\partial T}\right).
\end{equation}

 \end{subequations}
From above equations will get the black hole temperature as
\begin{equation}\label{ts1}
 T=\frac{1}{4\sqrt{\pi S}}+\frac{3 a \epsilon \pi^{\frac{3\epsilon}{2}}}{2 S^{\frac{3 \epsilon}{2}+1}}.
\end{equation}

 \begin{figure}
 \centering
\includegraphics[width=0.60\columnwidth]{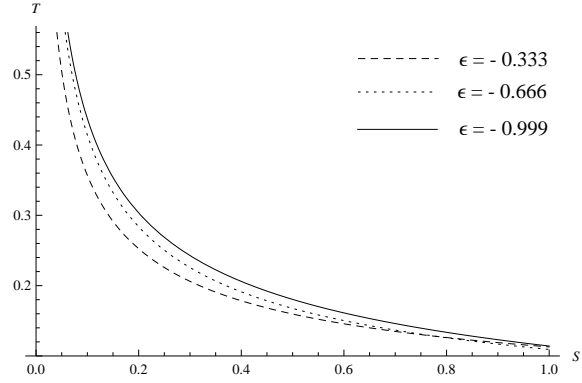}
\caption{Variation of temperature with entropy for different values of $\epsilon$, keeping $a=0.001$.}
\end{figure}
Fig(3) represents the T-S diagram.

Now we are looking for the heat capacity, $C$ of the black hole;
we get the heat capacity in terms of entropy and quintessence parameter as
\begin{equation}\label{cs}
 C=T \frac{\partial S}{\partial T}=- \frac{16 S^{3 \epsilon +5}+ 96 a \epsilon \pi^{\frac{3\epsilon +1}{2}} S^{\frac{3 \epsilon +9}{2}}}{8 S^{3 \epsilon +4}+144 a\epsilon^{2} \pi^{\frac{3\epsilon +1}{2}} S^{3 \epsilon +2}+ 96 a \epsilon \pi^{\frac{3\epsilon +1}{2}} S^{\frac{3 \epsilon +7}{2}}}.
\end{equation}
In Fig(4) we have plotted the variation of heat capacity for a fixed value of $a$ and for different values of
$\epsilon$, and the discontinuity in the figure pointing the phase transition behaviour.
\begin{figure}
 \centering
\includegraphics[width=0.60\columnwidth]{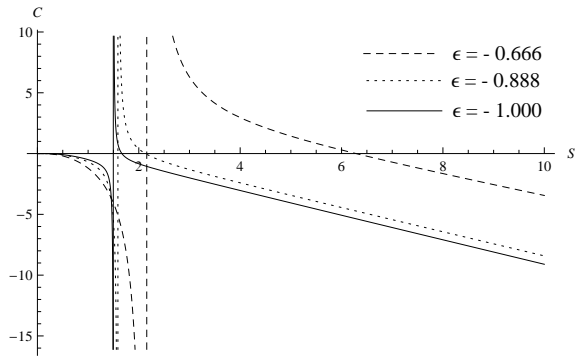}
\caption{Variation of heat capacity with entropy for different values of $\epsilon$, keeping $a=0.001$. }
\end{figure}

In the study of thermodynamics of the Schwarzschild black hole surrounded by quintessence we discussed the behaviour of temperature,
 mass and density of quintessence of the black hole with respect to its entropy. We further observed a discontinuity in the 
heat capacity which implies that the black hole undergoes a phase transition.

\section{Spectroscopy of Schwarzschild black hole surrounded by quintessence}

In this part of our work we discuss the properties of entropy quantization,
 the area and entropy spectrum.(In this section we are using $c=G=1$; and Planck length is written in units of $\hbar$).
We now find the the area spectrum by evaluating the adiabatic invariant integral\cite{Kunstatter}.

         The first law of black hole thermodynamics states\cite{Abbott}:

\begin{equation}\label{dm}
 dM=\frac{1}{4}T_{H}dA,
\end{equation}
where $M$ is the mass of the black hole, $T_{H}$ is the Hawking temperature and $A$ is the surface area of the black hole.

We can find the horizons if we fix the 
value of the state parameter of quintessence. In general,
\begin{equation}\label{sg}
 \kappa_{r}=\frac{1}{2}\left|\frac{df(r)}{dr}\right|_{r=particular~ horizon ~ radius},
\end{equation}
and for the present system of black hole it will be,
\begin{equation}\label{sgmr}
 \kappa_{r}(\epsilon)=\frac{1}{2}\left[\frac{2M}{r^{2}}+ \frac{a(3\epsilon+1)}{r^{3\epsilon+2}}\right].
\end{equation}
Hawking temperature is given by, 
\begin{equation}\label{tmr}
 T_{BH}=\frac{\hbar \kappa_{r} }{2 \pi}=\frac{\hbar}{2 \pi}\left[\frac{M}{r^{2}}+\frac{a(3\epsilon+1)}{2 r^{3\epsilon+2}}\right].
\end{equation}
Now we exclusively utilize the period of motion of outgoing wave, which is shown to be related to the vibrational frequency of the 
perturbed black hole, to quantize the area of the Schwarzschild black hole surrounded by quintessence.
It is well known that the gravity system in Kruskal coordinate is periodic with respect to the Euclidean time. Particles'
motion in this periodic gravity system also owns a period, which has been shown to be the inverse Hawking temperature\cite{Gibbons}.
To  find the area spectrum via the periodicity method, substitute (\ref{metric}) in K G equation,
\begin{equation}\label{kg}
 g^{\mu \nu}\partial_{\mu}\partial_{\nu}\Phi-\frac{m^{2}}{\hbar^{2}}\Phi=0.
\end{equation}
By adopting the wave equation ansatz for the scalar field we can get the solution of wave equation. On the other
hand we can also obtain the solution from the Hamilton-Jacobi equation.
\begin{equation}\label{hj}
  g^{\mu \nu}\partial_{\mu}S\partial_{\nu}S+m^{2}S=0,
\end{equation}
where $S$ is the action, and $S$ and $\Phi$ are related by,
\begin{equation}\label{phi}
 \Phi=exp\left[\frac{i}{\hbar}S(t,r,\theta,\phi)\right].
\end{equation}
For the spherically symmetric Schwarzschild black hole, the action can be decomposed as\cite{Damora,Srinivasan}
\begin{equation}\label{action}
 S(t,r,\theta,\phi)= -Et+ W(r)+ J(\theta,\phi),
\end{equation}
near the horizon J vanishes and W can be written as\cite{Damora,Srinivasan}
\begin{equation}\label{wr}
 W(r)=\frac{i \pi E}{f'(r_{H})},
\end{equation}
where we only consider the outgoing wave near the horizon. In this case, it is obvious that the wave function $\Phi$
outside the horizon can be expressed as the form,
\begin{equation}\label{phipsi}
 \Phi=exp\left[-\frac{i}{\hbar}Et\right] \psi(r_{H}),
\end{equation}

where 
\begin{equation}\label{psir}
 \psi(r_{H})=exp\left[-\frac{\pi E}{\hbar f'(r_{H})}\right],
\end{equation}
and from the above equation it is clear that $\Phi$ is a periodic function with period
\begin{equation}\label{tau}
 \tau=\frac{2 \pi \hbar}{E}.
\end{equation}

From(\ref{mr}) we can find
\begin{equation}\label{dmdrr}
 dM=\frac{1}{2}\left[1+\frac{3a\epsilon}{r^{3\epsilon+1}}\right]dr.
\end{equation}
Eliminating $M$ from (\ref{tmr}) using (\ref{mr}), we find
\begin{equation}\label{tr}
 T_{H}=\frac{\hbar}{4 \pi r}\left[1+\frac{3 a \epsilon}{r^{3\epsilon+1}}\right].
\end{equation}

Using the fact that the gravity system in Kruskal coordinate is periodic with respect to the Euclidean time, particles'
motion in this periodic gravity system also owns a period, which has been shown to be the inverse Hawking temperature.
Thus the relation between $\tau$ \& $T$ can be written as
\begin{equation}\label{taut}
 \tau=\frac{2 \pi}{\kappa_{r}}=\frac{\hbar}{T_{H}},
\end{equation}
hence
\begin{equation}\label{tk}
 T_{H}=\frac{\hbar \kappa_{r}}{2 \pi}.
\end{equation}
The expression of surface area of the event horizon is 
\begin{equation}\label{area}
 A=4 \pi r^{2},
\end{equation}
 and using (\ref{dmdrr})and (\ref{tk})
we get
\begin{equation}\label{deltaak}
 \Delta A=\frac{8 \pi dM}{\kappa_{r}}.
\end{equation}
According to the theory of Kunstatter\cite{Kunstatter}, the adiabatic invariant obeys the quantization rule: 
\begin{equation}\label{ai1}
 \int \frac{8 \pi dM}{\kappa_{r}} = n \hbar.
\end{equation}

It is not possible to study the area spectrum for general values of quintessence parameter. We choose three cases 
in which the quintessence parameter take the values $-1,-\frac{1}{3}$ and $-\frac{2}{3}$.
Among these $\epsilon=-\frac{1}{3}$ gives a Schwarzschild like case, $\epsilon=-1$ gives de-Sitter case and 
$\epsilon=-\frac{2}{3}$ is an intermediate case and we first discuss this case.
\subsection{Case 1; $\epsilon=-\frac{2}{3}$}
\label{sec:2}
\label{sec:2}
 In this case $f(r)$ becomes
\begin{equation}\label{fr1}
 f(r)=1-\frac{2M}{r}-ar.
\end{equation}
There are two horizons and they are given by
\begin{equation}\label{rb}
 r_{+}=\frac{1-\sqrt{1-8 a M}}{2 a},
\end{equation}
and
\begin{equation}\label{rc}
 r_{c}=\frac{1+\sqrt{1-8 a M}}{2 a}.
\end{equation}

Let us now specialize to a non-trivial case with a non-asymptotically flat space time called the near extremal case.
As for this case, the cosmological horizon $r_{c}$ is very close to the black hole horizon $r_{+}$. Hence one can make
 the following approximations:
\begin{equation}\label{rbrc1}
r_{+}\approx r_{c},      r_{+}+r_{c} \approx 2 r_{+},   r_{c}-r_{+}=\Delta r.
 \end{equation}
Now the surface gravity at $r=r_{+}$
\begin{equation}\label{kdr}
 \kappa_{+}=\frac{a \Delta r}{2 r_{+}}.
\end{equation}
 Using (\ref{taut}) and (\ref{tk}) and using the notion that  $M$ is the total mass of the black hole,
we can write,
\begin{equation}
 dM=\hbar \omega= 2 \pi T_{+}.
\end{equation}

Therefore, the adiabatic invariant can be rewritten as
\begin{equation}
 I=\int{\frac{dM}{\kappa_{+}}}=\frac{M}{\kappa_{+}}+c,
\end{equation}
where $c$ is a constant. Bohr-Sommerfeld quantization then implies that the mass spectrum is equally spaced, namely
\begin{equation}\label{mkb}
 M= n \hbar \kappa_{+}
\end{equation}
Now using  (\ref{rbrc1}), (\ref{kdr}) and (\ref{mkb}) we get
\begin{equation}\label{dmrk}
 \delta M=\frac{\Delta r\delta r_{+}}{16a r_{+}^{2}}=\hbar \kappa_{+}.
\end{equation}
On the other hand, the black hole horizon area is given by 
\begin{equation}
 A_{r_{+}}=4 \pi r_{+}^{2},
\end{equation}
Implementing the variation of the black hole horizon and using  (\ref{dmrk}), we have
\begin{equation}\label{deltaarea}
 \delta A_{r_{+}}=8 \pi r_{+} \delta r_{+}= 8 \pi\frac{16 a r_{+}^{3}\hbar \kappa_{+}}{\Delta r}.
\end{equation}
Substituting (\ref{rbrc1}) and (\ref{kdr}) into  (\ref{deltaarea}), one can obtain
\begin{equation}
 \delta A_{r_{+}}=16 \pi \hbar.
\end{equation}
Hence the area spectrum of a Schwarzschild black hole surrounded by quintessence, for the particular case of $\epsilon=-\frac{2}{3}$
is equal to
\begin{equation}
 A_{n}=16 \pi n \hbar.
\end{equation}
Now we can derive the entropy spectra using the definition of Bekenstein-Hawking entropy,
\begin{equation}
 S=\frac{A_{n}}{4 \hbar}=4 n \pi.
\end{equation}
Thus we have derived the area and entropy spectra of Schwarzschild black hole in the presence of
quintessence for a particular value of  $\epsilon=-\frac{2}{3}$. We find that the area and entropy spectra are equally spaced.

\subsection{Case 2; $\epsilon=-\frac{1}{3}$}
\label{sec:3}
In this case $f(r)$ will be
\begin{equation}
 f(r)=1-\frac{2M}{r}-a.
\end{equation}
It is a Schwarzschild like case, and we can follow the method used for Schwarzschild black hole.
The asymptotic quasinormal modes can be found, and it was found that the surface gravity term came in the adiabatic invariant.

The surface gravity is given by
\begin{equation}
 \kappa_{+}=\frac{(1-a)^{2}}{4M},
\end{equation}
 and thus the adiabatic invariant can be written as
\begin{equation}
 I=\int{\frac{dM}{\kappa_{+}}}=\int \frac{4M dM}{(1-a)^{2}}.
\end{equation}
Now integrating and giving the condition as $M=\frac{(1-a)r_{+}}{2}$ we get the area spectrum as \cite{Bekenstein,Maggiore}
\begin{equation}
 A_{n}= 8 \pi  n \hbar.
\end{equation}
 And hence the entropy spectrum is
\begin{equation}
 S=\frac{A_{n}}{4 \hbar}=2 n \pi.
\end{equation}

\subsection{Case 3; $\epsilon=-1$}
Here $\epsilon=-1$ and $f(r)$ will be
\begin{equation}
 f(r)=1-\frac{2M}{r}-a r^{2}.
\end{equation}

The surface gravity is given by
\begin{equation}
 \kappa_{+}=\frac{M}{r^{2}}-2ar.
\end{equation}

The adiabatic invariant of the system can be written as
\begin{equation}
 I=\int\frac{dM}{\kappa_{+}}=\frac{M}{\kappa_{+}}+c,
\end{equation}
where c is a constant.
And now the area spectrum is given by\cite{li}
\begin{equation}
 A_{n}=24 \pi  n \hbar,
\end{equation}
and the entropy spectrum is
\begin{equation}
 S=\frac{A_{n}}{4 \hbar}= 6 n \pi.
\end{equation}
 
\section{SUMMARY AND CONCLUSION}
We have investigated the effect of quintessence on the thermodynamic and spectroscopic properties of 
Schwarzschild black hole. We have derived the expressions for mass, density of quintessence, temperature and heat capacity
of the black hole surrounded by quintessence in terms of its entropy. Since entropy is the crucial factor which comes in the spectroscopic analysis, 
we could define the variations of the above given parameters in terms of entropy. And in the mass-entropy graph, the idea of area law is justified.
We have plotted the variation of quintessence density parameter with respect to entropy and found that density decreases with entropy.
Heat capacity of the black hole shows a discontinuity, which implies that the black hole may undergo a phase transition.  

In the spectroscopic analysis we have derived the adiabatic invariant integral using the periodicity method. 
We got the area and entropy spectra equidistant irrespective of the quintessence parameter but the spacing is different
for different values of state parameter. 
\section*{Acknowledgments}

TR wishes to thank UGC, New Delhi for financial support under RFSMS scheme.
VCK is thankful to CSIR, New Delhi for financial support under Emeritus Scientistship scheme and wishes
to acknowledge Associateship of IUCAA, Pune, India.


\end{document}